\documentstyle[pra,aps]{revtex}

\begin{document}

\title{Quantum fluctuations of mass for a mirror in vacuum}
\author{Marc-Thierry Jaekel $^a$ and Serge Reynaud $^b$}
\address{$(a)$ Laboratoire de Physique Th\'eorique de l'ENS \thanks{%
Unit\'e propre du CNRS associ\'ee \`a l'Ecole Normale Sup\'erieure et \`a %
l'Universit\'e Paris-Sud}, 24 rue Lhomond F75231 Paris Cedex 05 France\\
$(b)$ Laboratoire de Spectroscopie Hertzienne de l'ENS \thanks{%
Unit\'e de l'Ecole Normale Sup\'erieure et de l'Universit\'e Pierre et Marie
Curie associ\'ee du CNRS }, 4 place Jussieu F75252 Paris Cedex 05 France}
\date{{\sc Physics Letters} {\bf A 180} (1993) 9-14}

\maketitle


\def\beqn{\begin{equation}}
\def\eeqn{\end{equation}}
\def\beqa{\begin{eqnarray}}
\def\eeqa{\end{eqnarray}}



\def\q{\dot q}
\def\qq{\ddot{q}}
\def\dddq{\stackrel{...}{q}}
\def\dQ{\dot Q}
\def\dphi{\partial_t\phi}


\def\dt{\delta t}
\def\dx{\delta x}
\def\s{\sigma}
\def\xq{\xi}
\def\ds{\dot \s}
\def\dxq{\dot \xq}


\def\T{{\it T}}
\def\Tf{{\it T}}
\def\TfL{{\it T}^{(L)}}
\def\TfR{{\it T}^{(R)}}
\def\Ti{{\it T}_i}
\def\Te{{\it T}_*}
\def\Ts{t}
\def\ef{e}
\def\pf{p}
\def\efL{e^{(L)}}
\def\pfL{p^{(L)}}
\def\efR{e^{(R)}}
\def\pfR{p^{(R)}}
\def\ei{e_i}
\def\pin{p_i}
\def\eo{e_o}
\def\po{p_o}
\def\ee{e_*}
\def\pe{p_*}
\def\ec{e_c}
\def\pc{p_c}
\def\eiL{\ei^{(L)}}
\def\piL{\pin^{(L)}}
\def\eiR{\ei^{(R)}}
\def\piR{\pin^{(R)}}
\def\ecL{\ec^{(L)}}
\def\pcL{\pc^{(L)}}
\def\ecR{\ec^{(R)}}
\def\pcR{\pc^{(R)}}


\def\phif{\phi}
\def\psif{\psi}
\def\phii{\phi_i}
\def\psii{\psi_i}
\def\phio{\phi_o}
\def\psio{\psi_o}
\def\phie{\phi_*}
\def\psie{\psi_*}
\def\phic{\phi_c}
\def\psic{\psi_c}
\def\Phii{\Phi_i}
\def\Phic{\Phi_c}
\def\dvphi{\dot \varphi}
\def\dpsi{\dot \psi}


\def\ci{c_i}
\def\cc{c_c}
\def\CC{C_-}
\def\C+{C_+}

\def\S{S}
\def\O{\Omega}


\def\E{E}
\def\P{P}
\def\K{K}
\def\D{D}
\def\De{\Delta}
\def\Ef{E}
\def\Pf{P}
\def\Kf{K}
\def\Df{D}
\def\Def{\Delta}
\def\Ei{E_i}
\def\Pin{P_i}
\def\Ki{K_i}
\def\Di{D_i}
\def\Dei{\Delta_i}
\def\Ec{E_c}
\def\Pc{P_c}
\def\Kc{K_c}
\def\Dc{D_c}
\def\Fc{F_c}
\def\Gc{G_c}
\def\FcL{\Fc^{(L)}}
\def\GcL{\Gc^{(L)}}
\def\FcR{\Fc^{(R)}}
\def\GcR{\Gc^{(R)}}
\def\dphc{\Delta_c}
\def\kphc{\Gamma_c}
\def\Qc{Q_c}

\def\e{e}
\def\p{p}
\def\k{k}
\def\d{d}
\def\Des{\Delta_s}
\def\dph{\delta}


\def\t{\bar t}
\def\x{\bar x}
\def\bq{\bar q}
\def\bds{d\bar \sigma}
\def\bpartial{\bar \partial}
\def\bg{\bar g}
\def\bdS{d\bar S}

\def\bTf{\bar \Tf}
\def\bTfL{{\bar \Tf}^{(L)}}
\def\bTfR{{\bar \Tf}^{(R)}}
\def\bTi{\bar \Ti}
\def\bTs{\bar \Ts}

\def\bphi{\bar \phi}
\def\bpsi{\bar \psi}
\def\bphii{\bar \phii}
\def\bpsii{\bar \psii}
\def\bphio{\bar \phio}
\def\bpsio{\bar \psio}
\def\bPhi{\bar \Phi}

\def\bc{\bar c}
\def\bci{\bar \ci}
\def\bS{\bar S}

\def\bDef{\bar \Def}
\def\bDei{\bar \Dei}
\def\bDes{\bar \Des}
\def\be{\bar \e}
\def\bp{\bar \p}
\def\bk{\bar \k}
\def\bd{\bar \d}
\def\m{m}
\def\bF{\bar F}
\def\bG{\bar G}


\begin{abstract}
A mirror in vacuum is coupled to fluctuating quantum fields.
As a result, its energy-momentum and mass
fluctuate. We compute the correlation spectra of
force and mass fluctuations for a mirror at rest in vacuum (of a
scalar field in a two-dimensional space-time). The obtained
expressions agree with a mass correction equal to a vacuum
energy stored by the mirror.
We introduce a Lagrangian model which consistently
describes a scalar field coupled to a scatterer, with inertial
mass being a quantum variable.

\end{abstract}

\section{Introduction}

Quantum field fluctuations in vacuum lead to mechanical effects on
objects by which they are scattered \cite{Sciama}.
Two mirrors of a cavity scattering vacuum fields
feel a radiation pressure
(Casimir force), which results from
a difference of field energy densities between the inside
and outside of the cavity \cite{Casimir}.
Casimir energy can be seen
as a variation in the vacuum field energy stored in the cavity
\cite{JR1}. As known since Einstein \cite{E2},
any energy stored in a box modifies the inertial mass of the box.
A cavity moving in vacuum also feels an inertial force,
in agreement with Einstein's principle of inertia of energy
 \cite{JR8}.

Casimir energy is the length dependent part of the vacuum energy
stored by the cavity, i.e. the integral over frequency
of the mean energy density of vacuum fields multiplied
by the storage delay in the cavity \cite{JR1}.
Scattering by a partially transmitting mirror also involves a non
vanishing time delay which must lead to a stored
energy. From equivalence between energy and mass, vacuum
fluctuations are expected to induce a mass for a single mirror also.
For a pointlike mirror scattering a scalar field in a two
dimensional (2D) space-time, the relation between induced mass
($<\Delta m>$)
and time delay ($\tau$) takes the form \cite{JR1} (units are such
that light's velocity is equal to 1; $\hbar$ is Planck's constant):
\beqn
\label{m}
<\Delta m> =  \int_0^\infty {d\omega \over 2\pi}
\hbar \omega \tau[\omega]
\eeqn

The radiation pressure exerted on a mirror by a quantum field
 fluctuates, even in vacuum \cite{Barton,JR2}.
Total
energy and momentum remain conserved, whilst being continuously exchanged
between mirror and field through fluctuations. As a result of
energy-momentum conservation, an energy fluctuation absorbed by the mirror
changes its rest energy \cite{E1}. Hence, the mirror's mass also
exhibits fluctuations in vacuum. This can also be seen as a consequence
of energy storage (\ref{m}), since the energy density of vacuum
fields is itself a fluctuating quantity.

In this letter, we derive correlation spectra associated
with force and mass fluctuations for a mirror at rest in the vacuum of
a (2D) scalar field. We use two equivalent descriptions of a partially
transmitting mirror, in terms of a scattering matrix, and as a
pointlike source for the field.
The obtained spectra agree with
 expression (\ref{m}) for the mean value of the mass induced by vacuum
field fluctuations.
When evaluated for a mirror at rest, expression (\ref{m}) diverges. Actually,
neglecting the mirror's recoil during scattering can only be consistent
with an infinite mass. In order to remedy this defect, we
introduce a local relativistic Lagrangian which
takes recoil effects into account and entails conservation laws.
The mirror's mass then appears as a quantum variable.

\section{Force and mass fluctuations in vacuum}

For simplicity, we study the case of a pointlike mirror at rest (at
position $q$), which  scatters
a scalar field ($\phi$) in a (2D) space-time ($(x^\mu)_{\mu=0,1} = (t,x)$).
We first use a model previously introduced  for describing
partially transmitting mirrors  \cite{JR1,JR2}.
On each side of the mirror, the scalar field is the sum of
two freely propagating components:
\beqa
\label{mod}
\phi(t,x) &=& \varphi_{in}(t-x) + \psi_{out}(t+x) \qquad \qquad {\rm
for} \qquad
x < q\nonumber\\
\phi(t,x) &=& \varphi_{out}(t-x) + \psi_{in}(t+x) \qquad \qquad {\rm
for} \qquad
x > q
\eeqa
The outcoming fields are linearly determined from
the incoming fields by a scattering matrix (satisfying reality,
causality and unitarity conditions):
\beqn
\label{S}
\left(\matrix{\varphi_{out}[\omega]\cr
\psi_{out}[\omega]\cr}\right) = \S[\omega]
\left(\matrix{\varphi_{in}[\omega]\cr
\psi_{in}[\omega]\cr}\right) \qquad \qquad
\S[\omega] = \left(\matrix{s[\omega]&r[\omega]e^{-2i\omega q}\cr
r[\omega]e^{2i\omega q}&s[\omega]\cr}\right)
\eeqn
$s$ and $r$ are frequency dependent transmission and reflection
amplitudes.
We use the following notation for Fourier transforms:
$$f(t) =  \int_{-\infty}^\infty {d\omega \over 2\pi}  e^{-i\omega t}
f[\omega]$$
On both sides of the mirror, the field propagates freely and its
stress tensor ($\Tf^{\mu\nu}$) is divergenceless:
\beqa
\label{st}
\Tf^{00} &=& \Tf^{11} = {1\over2}(\partial_t \phi^2 +
\partial_x \phi^2)\nonumber\\
\Tf^{01} &=& \Tf^{10} = - \partial_t \phi
\partial_x \phi
\eeqa
The variation in time of the field
energy-momentum ($\Pf^\mu$)
is given by fluxes of the stress tensor through the
mirror (difference between left and right sides):
\beqa
\label{f}
\dot{\Pf}^0 &=& - F^0 = - \lbrace \dvphi_{in}(t-q)^2 -
\dpsi_{out}(t+q)^2\rbrace + \lbrace\dvphi_{out}(t-q)^2 -
 \dpsi_{in}(t+q)^2 \rbrace
\nonumber\\
\dot{\Pf}^1 &=& - F^1 = - \lbrace \dvphi_{in}(t-q)^2 +
\dpsi_{out}(t+q)^2\rbrace + \lbrace\dvphi_{out}(t-q)^2 +
 \dpsi_{in}(t+q)^2
\rbrace\nonumber\\
\eeqa
(at rest, $\q = 0$; a dot stands for time derivative).
In particular, the space component $F^1$ is the radiation pressure
felt by the mirror and has already been studied \cite{JR2}.
For a single mirror and vacuum input fields, its mean value
vanishes. However, because of
quantum field fluctuations, the force fluctuates and its correlations
do not vanish. The time component $F^0$ corresponds to an energy
exchange, and will appear in the following as responsible for mass
fluctuations.
In a vacuum input state, the two field components are uncorrelated
and have identical auto-correlations:
$$<\varphi_{in}[\omega]\varphi_{in}[\omega']> =
<\psi_{in}[\omega]\psi_{in}[\omega']> = {2\pi \over\omega^2}
\delta(\omega+\omega')\theta(\omega) {1\over2} \hbar \omega $$
($\theta$ is Heaviside's step function). Force fluctuations on a mirror at rest
are stationary:
$$<F^\mu(t)F^\nu(t')> - <F^\mu><F^\nu> = C_{F^\mu F^\nu}(t-t')$$
and are easily computed using Wick's rules
and scattering matrix unitarity \cite{JR2}:
\beqa
\label{ff}
C_{F^\mu F^\nu}[\omega] &=& 2 \hbar^2 \theta(\omega)
\int_0^\omega {d\omega' \over 2\pi}
\omega'(\omega-\omega')
\alpha^{\mu\nu}[\omega',\omega-\omega']\nonumber\\
\alpha^{00}[\omega,\omega']&=& Re\lbrace 1 - s[\omega]s[\omega'] -
r[\omega]r[\omega']\rbrace\nonumber\\
\alpha^{01}[\omega,\omega']&=& \alpha^{10}[\omega,\omega']=
0\nonumber\\
\alpha^{11}[\omega,\omega']&=& Re\lbrace 1 - s[\omega]s[\omega'] +
r[\omega]r[\omega']\rbrace
\eeqa
Correlation spectra in vacuum contain positive frequencies only
(field excitations necessarily have positive energies with respect to
vacuum). Force auto-correlations are positive and vanish at least as
$\omega^3$ around zero frequency (vacuum energy-momentum is conserved
on large time scales). Some properties still
distinguish the two components of the force. In the limit of perfect
reflection ($s = 0$, $r = -1$), fluctuations associated with the time
component vanish, while radiation pressure fluctuations subsist and
take a simple form \cite{JR2}:
\beqa
C_{F^0 F^0}[\omega] &=& 0 \nonumber\\
C_{F^1 F^1}[\omega] &=& {\hbar^2 \over 3 \pi} \theta(\omega) \omega^3\nonumber
\eeqa
Actually, for a perfect mirror, $F^0$ vanishes as an operator while
$F^1$ is twice the momentum density of the input field (see (\ref{f})).
In the general case and at low frequencies,
radiation pressure fluctuations
exhibit the same behaviour while
correlations of the time component actually behave
as $\omega^5$ (as a consequence of scattering matrix
unitarity, the first four coefficients in a
quasistatic expansion around zero frequency vanish):
\beqa
C_{F^0 F^0}[\omega] &\sim&
\lbrace (is'[0])^2 + (ir'[0])^2 \rbrace
 {\hbar^2 \over 12 \pi} \theta(\omega) \omega^5\nonumber\\
C_{F^1 F^1}[\omega] &\sim& r[0]^2 {\hbar^2 \over 3 \pi} \theta(\omega)
\omega^3\nonumber
\eeqa
Both energy and momentum of the field exhibit fluctuations. At the
quasistatic limit however, energy
fluctuations are much less important than momentum fluctuations.

By conservation of total energy-momentum, $F^\mu$ is equal to the
time derivative of the mirror's energy-momentum $\p^\mu$:
\beqn
\label{emb}
\dot{\p}^\mu =  F^\mu
\eeqn
As a consequence, the mirror's energy and momentum also fluctuate
and their correlation
spectra
are given by those of the force
(\ref{ff}):
$$C_{\p^\mu \p^\nu}[\omega] = {1\over \omega^2} C_{F^\mu F^\nu}[\omega]$$
For a mirror at rest, mass is equal to energy. Hence,
mass fluctuates and its correlations in vacuum are readily
obtained:
\beqn
\label{mm}
C_{mm}[\omega] = {1\over \omega^2} C_{F^0 F^0}[\omega]
\eeqn
At low frequency, mass fluctuations scale
as $\omega^3$, and the
mirror's mass can be considered as constant on a large time
scale.
The mass noise spectrum (which vanishes in
the limit of perfect reflection) contains only positive frequencies
in vacuum. Stationary correlations possessing this property correspond to
non commutative variables, so that
mass fluctuations have an irreducible quantum character.

\section{Mirror as a pointlike source}

The scattering model for partially transmitting mirrors (equations (\ref{mod})
and (\ref{S})) amounts to consider that
the mirror acts as a pointlike source (denoted by
$J$) for the scalar
field, which obeys equations of motion of the form:
\beqn
\label{fe}
(\partial_t^2-\partial_x^2)\phi(t,x)  - J(t) \delta(x-q) = 0
\eeqn
Different
types of coupling are described by different functions  for the
source $J$ in terms of the field $\phi$.
Equation (\ref{fe}) can be put in a form which expresses the
energy-momentum balance between mirror and field
($\Tf^{\mu\nu}$ is the field energy-momentum tensor (\ref{st})):
\beqa
\label{em}
\partial_\nu \Tf^{\mu\nu} =
- F^\mu  \delta(x-q)\qquad \qquad
F^\mu = - J  \partial^\mu \phi (q)
\eeqa
where $F^\mu$ is the force
exerted by the field on the mirror.
When $J$ depends linearly on the field $\phi$, force
is a quadratic form of the field and coupling through radiation
pressure is equivalent to the scattering model
(\ref{f}).
Equation (\ref{fe}) corresponds to continuous boundary conditions for the
field on the mirror, and
the scattering coefficients take the
following form:
\beqn
\label{r}
s[\omega] = 1 + r[\omega] \qquad \qquad
r[\omega] = - {\O[\omega] \over \O[\omega] - i\omega}
\eeqn
$\O$ is a causal positive function (analytic with $Re\O >0$ for
$Im\omega >0$), even and real on the real axis ($\O[\omega]^* =
\O[-\omega] = \O[\omega]$ for $\omega$ real),
and describes the linear response of the
source to the local field (field evaluated on the mirror):
$$J[\omega]= - 2 \O[\omega] \bphi[\omega] \qquad \qquad
\bphi(t) = \phi(t,q)$$
{}From now on, we only consider the simple model which corresponds to
instantaneous response of the source to the local field,
$J = - 2\O \bphi$ with $\O$ constant.
In this case, the force exerted by the field on the mirror
takes a simple form:
$$F^\mu =  \partial^\mu (\O \phi^2)(q)$$
showing (see (\ref{emb})) that there is a mass (energy at rest) correction
for the mirror which identifies with
a function of the local field:
\beqn
\label{lm}
\Delta m = \O\bphi^2
\eeqn
Positivity of the induced mass ($\O > 0$) can be seen as equivalent
to causality of field scattering (\ref{r}).

This simple model corresponds
to a lorentzian reflectivity ($\O$ playing
the role of a high
frequency reflection cut-off),
with a frequency
dependent reflection delay $\tau$ given by \cite{JR1}:
\beqn
\label{td}
det\S[\omega] = e^{i\Delta[\omega]} \qquad \qquad 2\tau[\omega] =
\partial_\omega \Delta[\omega] =
{2\O \over \O^2 + \omega^2}
\eeqn
($\Delta[\omega]$ is the sum of the two phase shifts at frequency
$\omega$).
The correlation spectrum ((\ref{mm}) and (\ref{ff})) can be rewritten
in terms of the reflection delay only:
\beqn
\label{sm}
C_{mm}[\omega] = 2\hbar^2 \theta(\omega)
\int_0^\omega {d\omega' \over 2\pi}
\omega'(\omega-\omega')\tau[\omega'] \tau[\omega-\omega']
\eeqn
Direct evaluation of the
correlations of $\Delta m$ using (\ref{lm}) leads to the  same
expression (the field is continuous on the mirror and its value given by
(\ref{mod}) and (\ref{S})):
$$C_{mm}(t) = 2 \O^2 C_{\bphi\bphi}(t)^2\qquad\qquad \O
C_{\bphi\bphi}[\omega] = \theta(\omega) \hbar \omega
\tau[\omega]$$
It has the form of
the correlation spectrum of
the square of a gaussian variable whose mean value is given by (\ref{m})
(which can also be obtained from (\ref{lm})).
The mean value and variance of the instantaneous
mass are simply related:
$$<\Delta m^2> - <\Delta m>^2 = C_{mm}(0) =
 2 <\Delta m>^2$$
This relation shows that mass has important fluctuations on short
time scales, although it remains practically constant in the low
frequency domain:
\beqn
\label{lf}
C_{mm}[\omega] \sim {\hbar^2 \over 6 \pi} \theta(\omega)
{\omega^3\over\O^2} \qquad \qquad {\rm for} \qquad \omega \ll \Omega
\eeqn
As expected, mass fluctuations vanish in
the limit of
 perfect reflection
($\tau=0$). For non zero reflection delay:
\beqn
\label{sm2}
C_{mm}[\omega] = {\hbar^2 \over 2\pi}\theta(\omega){\O^2 \over \omega
(1 + {\omega^2\over 4\O^2})}\lbrace
(1 + {\omega^2\over 2 \O^2})log(1 + {\omega^2 \over \O^2}) - {\omega\over\O}
tg^{-1}({\omega\over\O}) \rbrace
\eeqn

Coupling a mirror to vacuum fields by radiation pressure
modifies the dynamical nature of its rest energy or mass.
Through its dependence on the field, the mirror's mass
exhibits quantum fluctuations.
Its mean value is also modified (\ref{m}).
For non zero reflection delay this expression is infinite, and
diverges logarithmically (see (\ref{td})).
Actually, the  previous model
gives a description
which is consistent with momentum conservation only when the mirror has an
infinite mass (frequency of the scattered field is conserved
(\ref{S})). For a finite mass,
 recoil effects
become important for incoming fields with sufficiently high frequency.
 It is also known that radiation pressure fluctuations
make the mirror's position fluctuate \cite{JR5}. In next section, we
introduce a Lagrangian model which takes these effects into account.

\section{Lagrangian model of a pointlike scatterer}

We rewrite the quadratic coupling between
field and scatterer  in a relativistically
invariant form by introducing a model Lagrangian.
In a flat (2D) space-time, the
classical action for a scalar field
($\phi$) and a point mass $m$ (with time dependent position $q$),
can be written:
$${\it S}= \int d^2x {1\over 2} (\partial \phi)^2 - \int m
(1-\q^2)^{1\over 2} dt$$
which is equivalent to the
following Lagrangian density:
\beqn
\label{L}
{\it L} =
{1\over2} [\partial_t \phi^2 - \partial_x \phi^2]
- m (1-\q^2)^{1\over2} \delta(x-q)
\eeqn
In accordance with the previous discussion, we consider that the scatterer's
mass depends on the field:
\beqn
\label{mb}
m = m_b + \Delta m
\eeqn
where $m_b$
is a "bare" mass and $\Delta m$ is a
(positive) scalar function of the local field ($\bphi$).
$\Delta m$ also represents the interaction of the
field with
sources located on the scatterer.
Quadratic coupling corresponds to $\Delta m = \O \bphi^2$,
$(\O > 0)$.

 The resulting Euler-Lagrange equations
are found to provide the field equations ((\ref{fe}) and
(\ref{em})) with:
$$J = - (1-\q^2)^{1\over 2} {\delta m\over \delta \bphi}$$
They also provide the equation of motion for the scatterer's position:
\beqn
\label{emm}
{d \over dt}({m \q \over (1-\q^2)^{1\over 2}}) = F^1
\eeqn
The latter coincides with the usual equation of motion
of a massive particle submitted to a force. However,
it involves a mass which is no longer constant, but
depends on the coupled field and on the particle's position.
Total energy-momentum is derived from Lagrangian (\ref{L}) and is the
sum of the field energy-momentum (given by (\ref{st})) and of the
scatterer's energy-momentum (there is no additional interaction term)
given by:
\beqn
\label{ep}
\p^0 = {m \over (1-\q^2)^{1\over 2}}
\qquad \qquad \p^1 = {m \q \over (1-\q^2)^{1\over 2}} \qquad \qquad
(\p^0)^2 = (\p^1)^2 + m^2
\eeqn
Energy-momentum conservation is obeyed (\ref{emb}).
It follows that
recoil is now included in the treatment of coupled mirror and field.
The following inequality results from (\ref{ep}):
$$|\p^1| \le |\p^0| \qquad \qquad {\rm i.e.} \qquad \qquad |\q| \le 1$$
The scatterer's velocity always remains smaller than light's
velocity. The position of the scatterer follows
a relativistic
stochastic process which respects causality.

The scatterer's mass is no longer a parameter and must be treated as
a quantum operator in the equation of motion (\ref{emm}). Mass
exhibits quantum fluctuations due to its dependence on the scattered
field. Furthermore, because of its dependence on the scatterer's position
it does not commute with
the scatterer's momentum or velocity.
When neglecting the quantum fluctuations of position,
the correlation spectrum of mass fluctuations
is given by (\ref{sm}) or (\ref{sm2}).
This spectrum gives an approximation to
mass fluctuations in the frequency domain where
recoil effects can be neglected ($\hbar \omega \ll <m>$).
For low frequencies (smaller than the reflection cut-off $\O$),
mass is practically constant (\ref{lf}), and
equation of motion (\ref{emm}) is well approximated by the
Newton law. Then, linear response formalism provides a
consistent treatment of quantum fluctations of field and
scatterer's position \cite{JR5}. However, for higher frequencies
($\omega \sim \O$),
mass can no longer be considered as constant in (\ref{emm}), and its
fluctuations must be taken into account.
For frequencies comparable with the scatterer's mass ($\hbar \omega
\sim <m>$), recoil effects
 must be included. In particular, the scattering matrix differs from
(\ref{S}) for high frequencies. High frequencies contribute
significantly to stored energy (\ref{m}), so that recoil effects
must be accounted for when determining the induced mass.

\section{Discussion}

Coupling a scatterer to quantum fields through radiation pressure
results in  modifications of its mass.
A single mirror moving
in vacuum experiences a motional force
which contains an inertial term. The induced mass is
the energy
stored on vacuum fields due to scattering time delay,
in conformity with the situation for a cavity \cite{JR8}. In
a cavity built from partially transmitting mirrors,
storage times depend on the
roundtrip time within the cavity and reflection
delays on the mirrors. Both terms are known to contribute to Casimir
energy \cite{JR1}. Similarly, both terms contribute to the
inertial force
acting on the cavity as a whole. This means that vacuum fluctuations
act as a Lorentz invariant
source of inertia (in accordance with the principle of inertia of
energy), not only for field energy stored within the cavity
\cite{JR8}, but also for the energy associated with reflection delays.

Motion in vacuum cannot remain causal and stable when the
mass induced by field fluctuations becomes infinite \cite{Rohrlich}.
As recoil and position fluctuations of the scatterer
significantly influence field scattering
at high frequencies,
the energy stored on vacuum fields is expected to remain finite and
smaller than the quasistatic mass (i.e $m_b > 0$ in (\ref{mb})).
Then, motion caused by a force including radiative reaction of
quantum fields can be shown to be described by a passive and therefore causal
response function \cite{JR4}.
Mass generation through vacuum energy storage
appears as a consistent way to entail
causality and stability of motion in vacuum.
Vacuum fluctuations are also known to contribute to
the electron self energy. This must lead to fluctuations of the
electromagnetic mass of the electron, that could be involved in a
solution of the stability problem of electron motion
\cite{Rohrlich}.

Mass arising from vacuum fluctuations is a quantum variable.
Mass fluctuations vanish on large time scales so that
mass measured in the low frequency domain appears as a
constant, as it should. On shorter time scales however (comparable
with reflection delay),
fluctuations of inertial mass become significant.
{}From the principle of
equivalence, gravitational mass is expected
to exhibit the same fluctuations \cite{Greenberger}, thus backing up
the necessity
that gravitation also has quantum fluctuations \cite{de Witt}.

\begin{flushleft}
{\bf Acknowledgements}
\end{flushleft}
We acknowledge discussions with J.M. Courty, A. Heidmann, P.A. Maia Neto, J.
Maillard and correspondence with G. Barton.

\end{document}